\documentclass[fleqn,useAMS]{mnras}
\usepackage{graphicx}	
\usepackage{amsmath}	
\usepackage{amssymb}	
\usepackage{multicol}        
\usepackage{bm}		
\usepackage{pdflscape}	

\usepackage[T1]{fontenc}
\usepackage{ae,aecompl}
\usepackage{booktabs}
\usepackage{footnote}

\usepackage[linewidth=1pt]{mdframed}
\usepackage{lipsum}

\DeclareRobustCommand{\VAN}[3]{#2}
\let\VANthebibliography\thebibliography
\def\thebibliography{\DeclareRobustCommand{\VAN}[3]{##3}\VANthebibliography}
\title{On the azimuthal alignment of quasars spin vector in large quasar groups and cosmic strings}
\author[R. Slagter et al.]{Reinoud J. Slagter,$^{1}$\thanks{E-mail: info@asfyon.com}
Pieter G. Miedema,$^{2}$\thanks{E-mail: pietermiedema@gmail.com}
\\
$^{1}$ASFYON, Astronomisch Fysisch Onderzoek Nederland \\University of Amsterdam, The Netherlands.\\
$^{2}$ASFYON, Astronomisch Fysisch Onderzoek Nederland}
\date{Accepted XXX. Received YYY; in original form ZZZ}
\pubyear{2020}
\begin{document}
\label{firstpage}
\pagerange{\pageref{firstpage}--\pageref{lastpage}}
\maketitle
\begin{abstract}
We find evidence of the alignment of the azimuthal angle of the spin vectors of quasars in their host galaxy in large quasar groups of different redshift. This effect could be explained by symmetry breaking of the scalar-gauge field of  cosmic strings in the early universe. It is expected that this effect will be more profound for higher red shift.
\end{abstract}
\section{Introduction}
Quasars are the most extensively investigated objects in astronomy. Astronomers are now convinced that a quasar is nothing but an active galactic nuclei (AGN) and consists of a violent eruption of radiation (in the optical as well as in the radio range), initiated by a spinning black hole (Kerr black hole) surrounded by an accretion disk\footnote{The book of D'Onofrio, "Fifty Years of Quasars" (D'Onofrio, et al., 2012) presents a clear overview.}. Quasars (and so their host galaxies) frequently appear in groups, i.e.,  large quasar groups (LQG).

There is observational evidence that in rich LQG the spin vector of the quasars are correlated, i.e., aligned in preferred azimuthal and polar angles.  Observational evidence of the correlation of spin vectors on Mpc scale of compact object such as quasars, goes back decades.  Quite recently, there was new  observational evidence for large-scale alignment of quasar optical polarization vectors in LQG (Hutsemaekers, et al., 2005, 2014). See also  the clear overview of Pelgrims and references therein. (Pelgrims, 2016). The same presence of large-scale spatial coherence was found by Taylor, et al. (Taylor, et al., 2016) by studying the distribution of radio jet position angles of radio galaxies over an area of one square degree in the ELAIS N1 field. The same conclusion was found by Pelgrims, et al., (Pelgrims, et al., 2016)

This mysterious coherence cannot be explained by mutual interaction.  
It is conjectured, that the explanation  is of  cosmological origin.
Several attempts were made to explain this phenomenon of cosmological origin. The alignment could be caused by primordial magnetic fields seeded by cosmic string loops (Poltis, et al., 2010). Axion-photon mixing in external magnetic fields is also considered as possible explanation (Payez, et al., 2008). 
Another explation could be delivered by the effects of pseudoscalar-photon mixing on electromagnetic radiation in the presence of correlated extragalactic magnetic fields. When one models the Universe as a collection of magnetic domains and study the propagation of radiation through them, then the correlations between Stokes parameters over large scales  consistently explains the observed large-scale alignment of quasar polarizations at different redshifts within the framework of the big bang model (Agarwal, et al., 2011).
However, one must realize that the correlation of the position angle alone is not enough to  prove the alignment of the quasar's spin vector. One needs the three dimensional orientation, i.e., the azimuthal and polar angles.
Investigations of the alignment of spin vectors of galaxies in clusters  (in the Sloan Digital Sky Survey, SDSS), show some anisotropic distribution (Flin, et al., 1986; Aryal, et al., 2008; Yadav, et al., 2016).
However, in large clusters it was found  that there is isotropy (Yadav, 2017, Malla, et al., 2019). This will confirm the conjecture that in smaller groups of quasars this alignment will be found and  explained  cosmologically (Slagter, 2016, 2017, 2018).

In this paper we propose an explanation of the alignment of the azimuthal angle of the spin vectors of groups of quasars (in their host galaxy). This alignment depends on the redshift and the number of quasars in the quasar group. 
In our general relativistic theoretical model, the same azimuthal angle appears  in the solution of the field equations as  trigonometric  functions. So observations of the spin vectors can be directly compared with the theoretical prediction. 
The recent observations of already mature galaxies at the early  epoch of the universe, i.e., when the cosmos was less than 7 percent of its present age of 13.7 billion years, support the  viewpoint that phenomena such as the alignment properties of quasar groups, emerged  at the early stages of our universe.

The observation of the structure and objects in our universe at the present epoch, are without doubt related to fundamental processes at the very early stage of that universe, i.e., the Planck scale of $\sim 10^{19}$ GeV. It is believed that the universe underwent violent phase-transition during that epoch. 
One of these processes was the famous Brout-Englert-Higgs (BEH) mechanism. 
Breaking of the initial symmetry leads to the mass spectrum we observe now. The Higgs particle (or Higgs field), responsible for this breaking, was recently observed at CERN with a mass of $\sim 125$ GeV. 

The symmetry breaking can be mathematically elegantly formulated in terms of Lie groups of quantum-chromo-dynamics (SU(3)) and quantum-electro-dynamics (SU(2)$\otimes$ U(1) of the electroweak unification). It is conjectured that even at high enough energy, these groups are sub-groups in a grant unified theory (GUT), first proposed by Georgi and Glashow.   
It is the hope of many physicists that eventually a quantum-gravity model will emerge, where Einstein's general relativity theory will be unified with quantum mechanics.
The features of a black hole, for example, can only be described by a quantum gravity model.
Many solutions are possible when the Higgs field, the basis of our Standard Model of particle physics, will be incorporated in general relativity theory (GRT). In fact, this scalar field was already necessary as the order parameter in the theory of superconductivity: the famous and experimental demonstrated Meissner effect. For a nice overview, see Felsager (Felsager, 1987) and Manton, et al. (Manton, et al., 2007).
\begin{figure}
\centerline{\includegraphics[scale=.40]{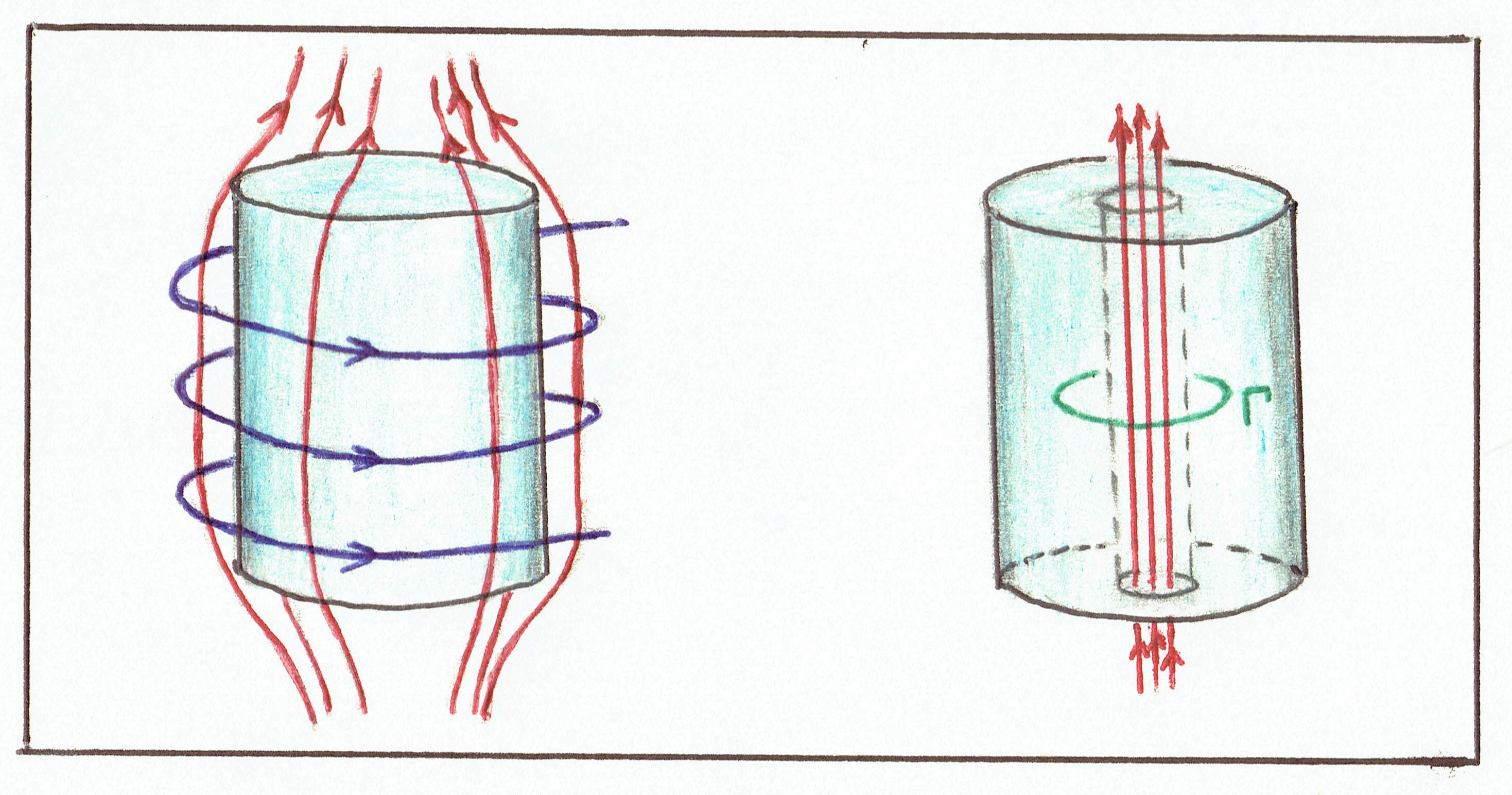}}
\centerline{\includegraphics[scale=.40]{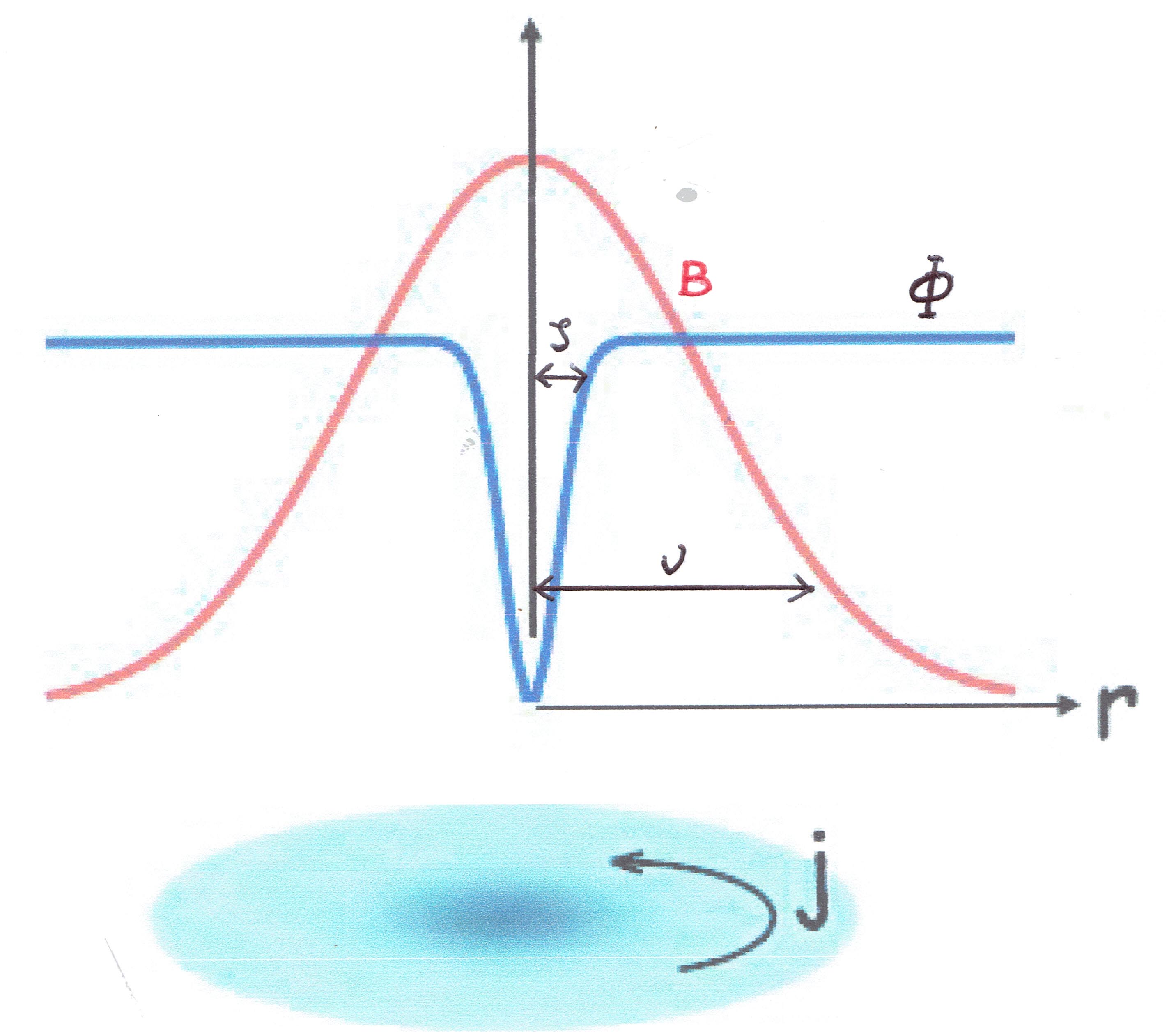}}
\centerline{\includegraphics[scale=.45]{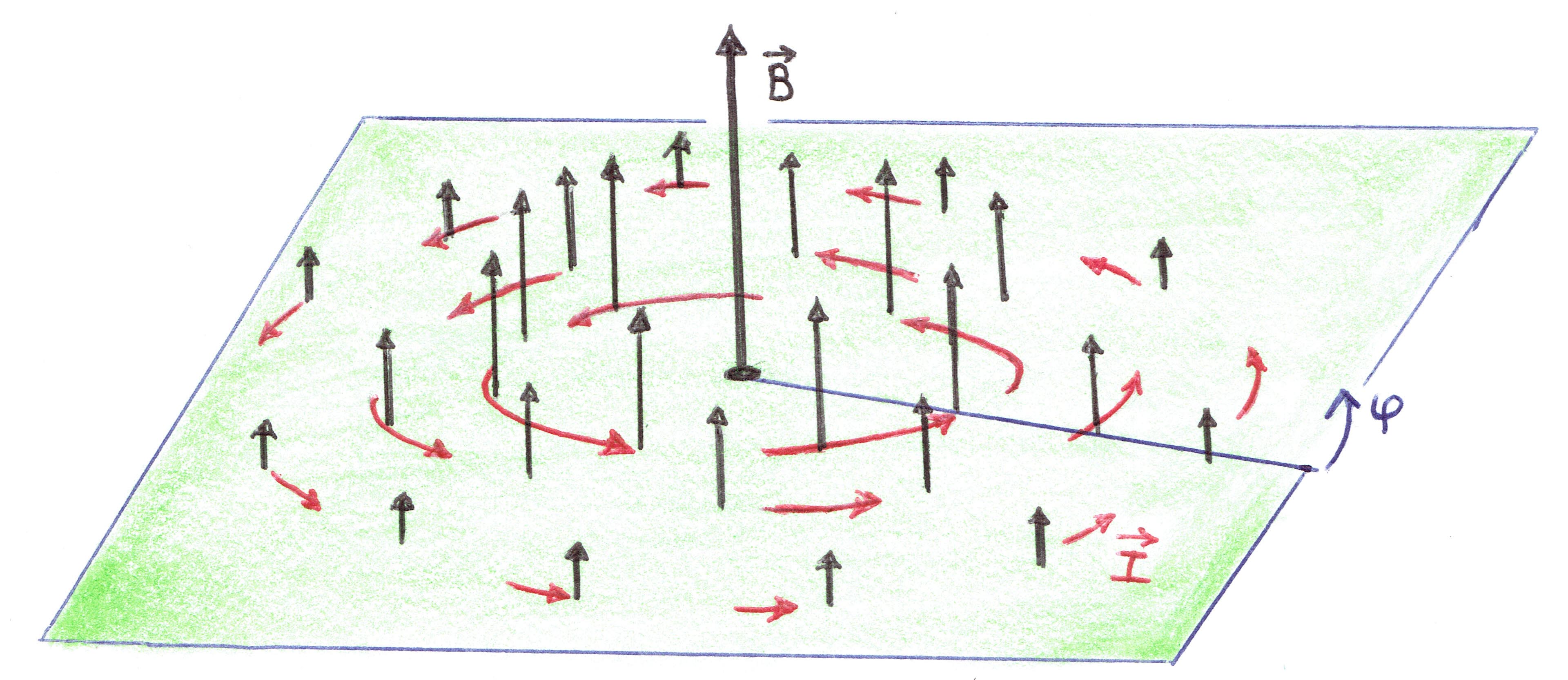}}
\caption{{\it The formation of quantized vortices. Abrikosov vortex in mixed state of quantized flux lines. Vortex supercurrents are sketched by round arrows in red. The radial dependence of the order parameter $\Phi$ is sketched as well as the magnetic field B. In type-II superconductivity the coherence length ($\zeta$) is much smaller than the penetration length ($\nu$).}}
\end{figure} 
If one places a metal cylinder (or ring) in an external magnetic field and one decreases the temperature below a certain critical temperature, then the magnetic field is expelled from the cylinder. A current is induced in the outer layers of the cylinder, which prevent the magnetic field from penetrating into the outer layers. 
The magnetic field is described by a gauge potential ${\bf A}$. It is this potential which penetrates into the metal and can change the phases of electrons passing by. When one removes the external magnetic field, then a part of the magnetic field lines is trapped by the surface current. The ring is transformed into a superconducting solenoid.

\begin{figure}
\centerline{\includegraphics[scale=.5]{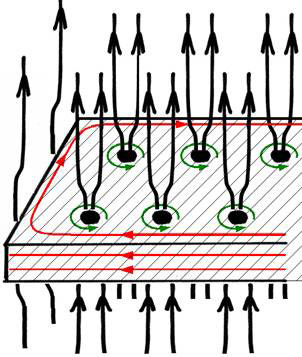}}
\centerline{\includegraphics[scale=.45]{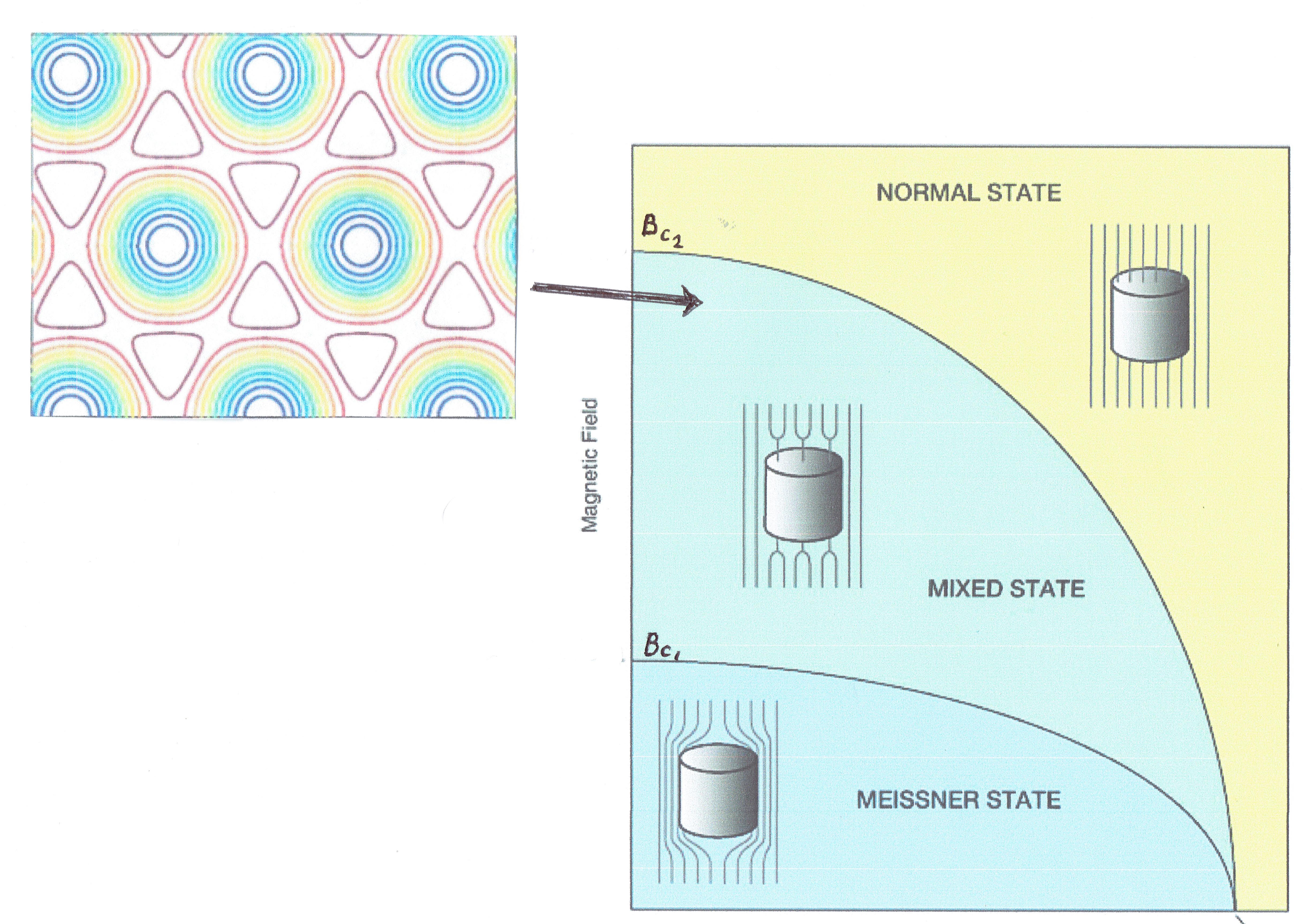}}
\caption{{\it Left: stable hexagonal Abrikosov lattice in mixed state of quantized fluxes lines. Vortex supercurrents are sketched by round arrows (green). Below: closely packed vortices near the critical region $B_{c_2}$}}
\end{figure}
In the superconducting state, there are Cooper pairs of electrons, which act like bosons, while the electron is a fermion. This "trapping" of magnetic flux happens if the temperature decreases. The superconducting state also depends on the strength of the external magnetic field. If we twist, for example, a solenoid  around the superconducting ring and  increase the magnetic field, thin quantized vortices are formed where the normal state of the metal is reestablished. This effect was first observed by Abrikosov (Abrikosov,1957) and is described by the famous Ginzburg-Landau (GL) equations (Ginsberg, et al., 1950). See figure 1 and 2.

The same Higgs field enters also in the model of inflationary cosmology, a phase of the universe, where the expansion is described by  an exponential function of the cosmological time. This model can explain some obstinate problems in the standard Friedmann-Lema\^itre-Robertson-Walker (FLRW) cosmological model, such as the horizon problem.

During the GUT phase-transitions a host of exotic objects may have formed, such as monopoles, domain walls and cosmic strings (CS).  
They are called  topological defects and are fully characterized by the scalar-gauge field ($\Phi,{\bf A}$).
Nielsen and Olesen (Nielsen, et al.,1973), were the first who formulated the relativistic string-like object in the Abelian Higgs model (also called Nambu strings). They are cylindrical symmetric and topological stable. 
In this model, which can be extended to Einstein's gravity theory (Garfinkle, 1985), the complex scalar field is written in polar coordinates as ${\bf \Phi}=\Phi(t,r)e^{in\varphi}$, where the phase contains the azimuthal angle and $n$ is the winding number (topological charge). It determines the phase jump $2\pi n$ when the Higgs field makes a closed curve around the string-like configuration. 
It is believed that only cosmic strings could survive the rapid expansion of the universe during the inflationary epoch.
However,  up until today, no experimental evidence of these objects are found and recent measurements of the microwave background power spectrum from COBE and WAMP show that cosmic strings could not provide an adequate explanation for the bulk of density perturbations. Gravitational waves detection will further put stringent bounds on these cosmic strings. 
Yet it is conjectured that in any field theory which admits cosmic strings, a network of strings inevitable forms at some point during the early universe and persists to the present time (Vilenkin,1994). 

New boost  to the field of cosmic strings emerged when it was realized that in string theory (or M-theory) super-massive CS  must be formed at an energy scale  much higher than the GUT scale. At this scale, the gravitational impact is high, because the CS builds up a huge mass in the bulk spacetime (Randall, et al. 1999). The warpfactor (or scale factor) will enter the  field equations and causes an amplification of the first and second order perturbations of the field variables (Slagter, 2016, 2017, 2018).   

This manuscript is organized as follows.
In section 2 we summarize the theoretical model. In section 3 we depict for six LQG the azimuthal angle of the spin vector and made the connection with the predictions of the theoretical model.
\section{\label{sec:level2}Summary of the model}
When gravity comes into play at very small scales and during the phase transition, one can write the FLRW spacetime as 
\begin{eqnarray}
ds^2={\cal W}(t,r,y)^2\Bigl[e^{2(\gamma(t,r)-\psi(t,r))}(-dt^2+ dr^2)+e^{2\psi(t,r)}dz^2\cr
+r^2 e^{-2\psi(t,r)}d\varphi^2\Bigr]+ dy^2,\label{eqn1}
\end{eqnarray}
where $y$ represents the extra dimension, ${\cal W}(t,r,y)$ the warpfactor or dilaton field (Slagter, et al., 2016). 
\begin{figure}
\centerline{\includegraphics[scale=.58]{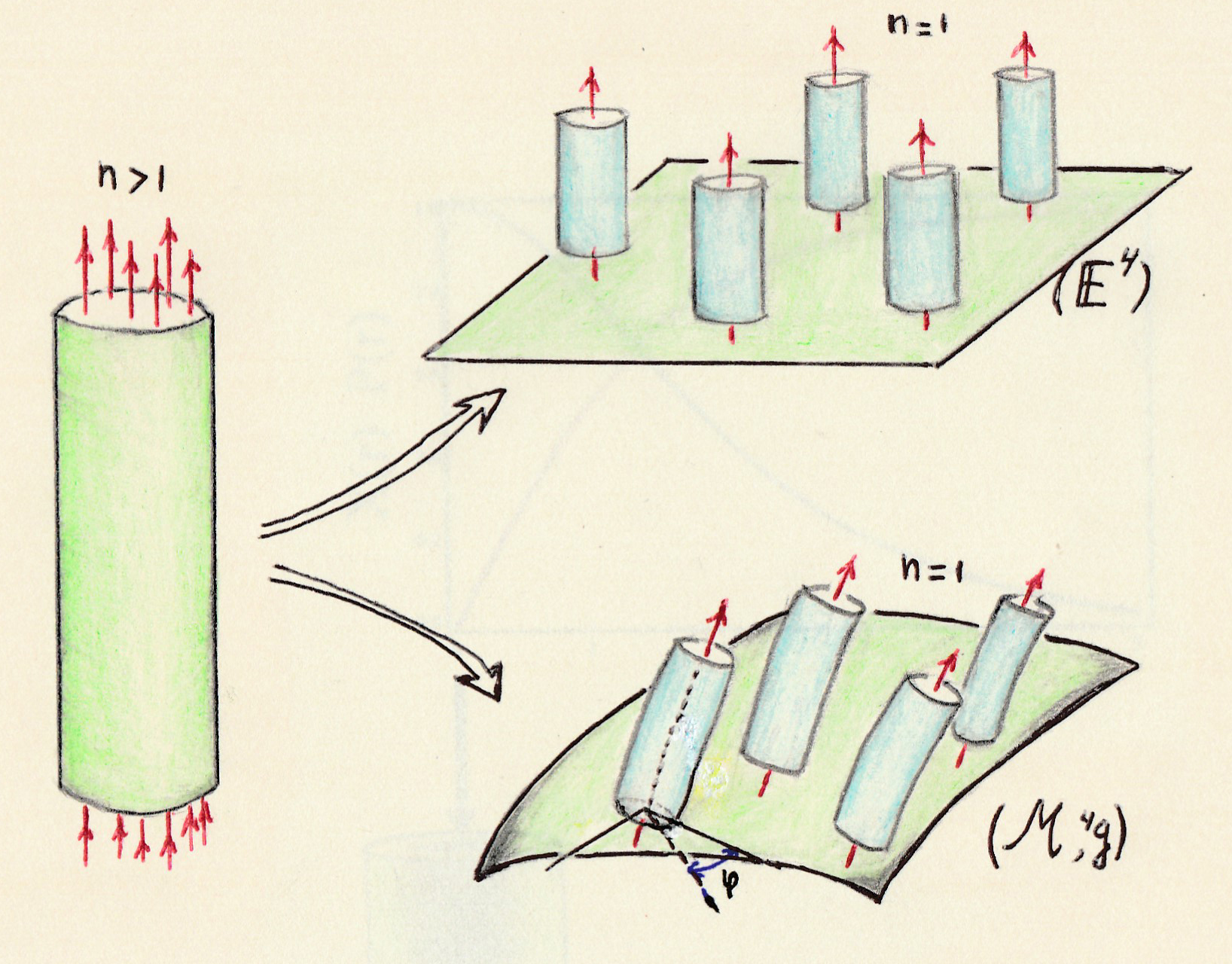}}
\caption{{\it Excitation and decay of a high multiplicity vortex into correlated vortices of unit flux $n=1$. Top: the Abrikosov lattice in Euclidean space. Bottom: correlated vortices with preferred azimuthal angle $\varphi$ in curved spacetime after the symmetry breaking. }}
\end{figure}

The complex scalar (Higgs) field and gauge field $A_\mu$ (Maxwell field) are written as
\begin{equation} 
{\bf \Phi}=\Phi(t,r)e^{in\varphi},\qquad A_\mu =\frac{n}{e}\bigl[P(t, r)-1\bigr]\nabla_\mu\varphi,\label{eqn2}
\end{equation}
where $n$ also determines  the magnetic flux of the vortex, which is quantized by $ \frac{2\pi n}{e}$.

If one writes out the field equations, then the azimuthal angle $\varphi$ will of
course not enter the PDE's, because the model is axially symmetric.
If the axial symmetry is dynamically broken, an off-diagonal metric function will appear, i.e., $g_{t\varphi}$. The spacetime will then possess 2 in stead if 3 Killing vectors. Quantum fluctuations will then excite the vortex. 

It is remarkable that this symmetry breaking is comparable with the phase transition in type II superconductivity, considered in our model.
Self-gravitating objects in GRT in equilibrium exhibit also analogue with the mathematical model of the MacLaurin-Jacobi sequences and its bifurcation points (Lebovitz, 1967).

After the excitation, the vortex configuration returns to its original axially symmetric situation, but an imprint will be left over in the azimuthal dependency of the orientations of the clustering of Abrikosov vortices lattice. It is caused by the fact that the energy of the vortex is proportional with $n^2$. So there can be no exact ground state for the string carrying multiple flux quanta. For $n$=1 we have the minimal energy situation, which is stable as it cannot decay into topological trivial field.
See figure 3. The topological charge can also be seen as the net number of new type of particles.

The excitation can be best described in an approximation scheme, were we expand the field variables as
\begin{eqnarray}
g_{\mu\nu}=\bar g_{\mu\nu}({\bf x})+ \frac{1}{\omega}h_{\mu\nu}({\bf x},\xi)+\frac{1}{\omega^2}k_{\mu\nu}({\bf x},\xi) + \cdots, \cr
A_\mu=\bar A_\mu ({\bf x})+\frac{1}{\omega}B_\mu ({\bf x},\xi) +\frac{1}{\omega^2}C_\mu ({\bf x},\xi) +\cdots ,\cr
\Phi=\bar\Phi({\bf x}) +\frac{1}{\omega}\Psi({\bf x}, \xi)+\frac{1}{\omega^2}\Xi({\bf x}, \xi)+\cdots,\label{eqn3}\qquad
\end{eqnarray}
where we write the subsequent orders of the scalar field as (Slagter, 2018)
\begin{eqnarray}
\bar\Phi =\eta \bar X(t,r) e^{i n_1 \varphi},\label{eqn4}\\ \Psi = Y(t,r,\xi) e^{i n_2 \varphi},\label{eqn5}\\\Xi = Z(t,r,\xi) e^{i n_3 \varphi}\label{eqn6},
\end{eqnarray}
$1/\omega$ represents the expansion parameter in the so-called multiple-scale approximation(Choquet-Brruhat, 1969, 1977; Slagter, 2001.

The relevant energy-momentum tensor components are
\begin{eqnarray}
{^{4}T}_{t\varphi}^{(0)}=\bar X\bar P\dot Y n_1{\bf sin}[(n_2-n_1)\varphi],\label{eqn7}
\end{eqnarray}
\begin{eqnarray}
{^{4}T}_{\varphi\varphi}^{(0)}=e^{-2\gamma}r^2\dot Y(\partial_t\bar X-\partial_r\bar X){\bf cos}[(n_2-n_1)\varphi]\cr 
+\frac{n_1 e^{2\bar\psi-2\bar\gamma}}{\bar W_1^2 e}\dot B(\partial_r\bar P-\partial_t\bar P),\label{eqn8}
\end{eqnarray}
\begin{eqnarray}
{^{4}T}_{tt}^{(0)}=\dot Y^2+\dot Y(\partial_t\bar X+\partial_r \bar X){\bf cos} [(n_2-n_1)\varphi ]\cr
+\frac{e^{2\bar\psi}}{\bar W_1^2 r^2 e}\Bigr(e \dot B^2+n_1\dot B(\partial_r\bar P+\partial_t\bar P)\Bigr),\label{eqn9}
\end{eqnarray}
while the background term ${^{4}\bar T}_{t\varphi}=0$. We conclude   that the axially symmetry is broken already to first order: the azimuthal angle ($\varphi$) dependency appears in the first and second order terms. In the expression for  ${^{4}T}_{tt}^{(0)}$ we observe that the scale factor $W_1$ enters the denominator. So if the scale increases, the contribution of the ${\bf cos} [(n_2-n_1)\varphi ]$ will become dominant. In the second order terms there appear terms like $\cos(n_3-n_2)\varphi$ (Slagter, 2017). These terms have extrema which differ mod$(\frac{\pi}{n})$. 
After the excitation of the vortex with multiplicity $n$, it will decay into $n$ vortices of unit flux in a regular lattice (just as the Abrikosov vortices form a hexagonal lattice such that the energy is minimal). 
The calculation of  the forces between the vortices, is complicated by the gravitational contribution. In the Bogomol'nyi (Bogomol'nyi, 1976) approximation, where the masses of the Higgs  and gauge particles are equal, one proves that there is equilibrium. In general, one must solve the time dependent GL equations, which can only be done numerically.
From the expression of ${^{4}T}_{t\varphi}^{(0)}$, we conclude that when the configuration returns to its original ground state and $n_2=n_1=1$ ($\sin(n_2-n_1)\varphi\rightarrow 0$)  and the axially symmetry is restored to first order. The term $\cos(n_2-n_1\varphi)$ in ${^{4}T}_{\varphi\varphi}^{(0)}$ has its maximum. So there is an emergent imprint of a preferred azimuthal angle $\varphi$ on the lattice of vortices when the ground state is reached ($n$=1).
\section{\label{sec:level2a}The quasar connection}
We gathered quasar data from the NASA/IPAC extragallactic database (NED) and SIMBAD database. We investigated 4 LQG (Park, et al., 2015) with a average redshift of $z=1.55, 1.51$, $1.06$ and $0.74$. We extract the position angle ($p$) and eccentricity ($\epsilon$) of the host galaxy. The data are based on the assumption that $H_o=73.0, \Omega_M=0.27$ and $\Omega_{vac}=0.73$
\begin{figure}
\centerline{\includegraphics[scale=.58]{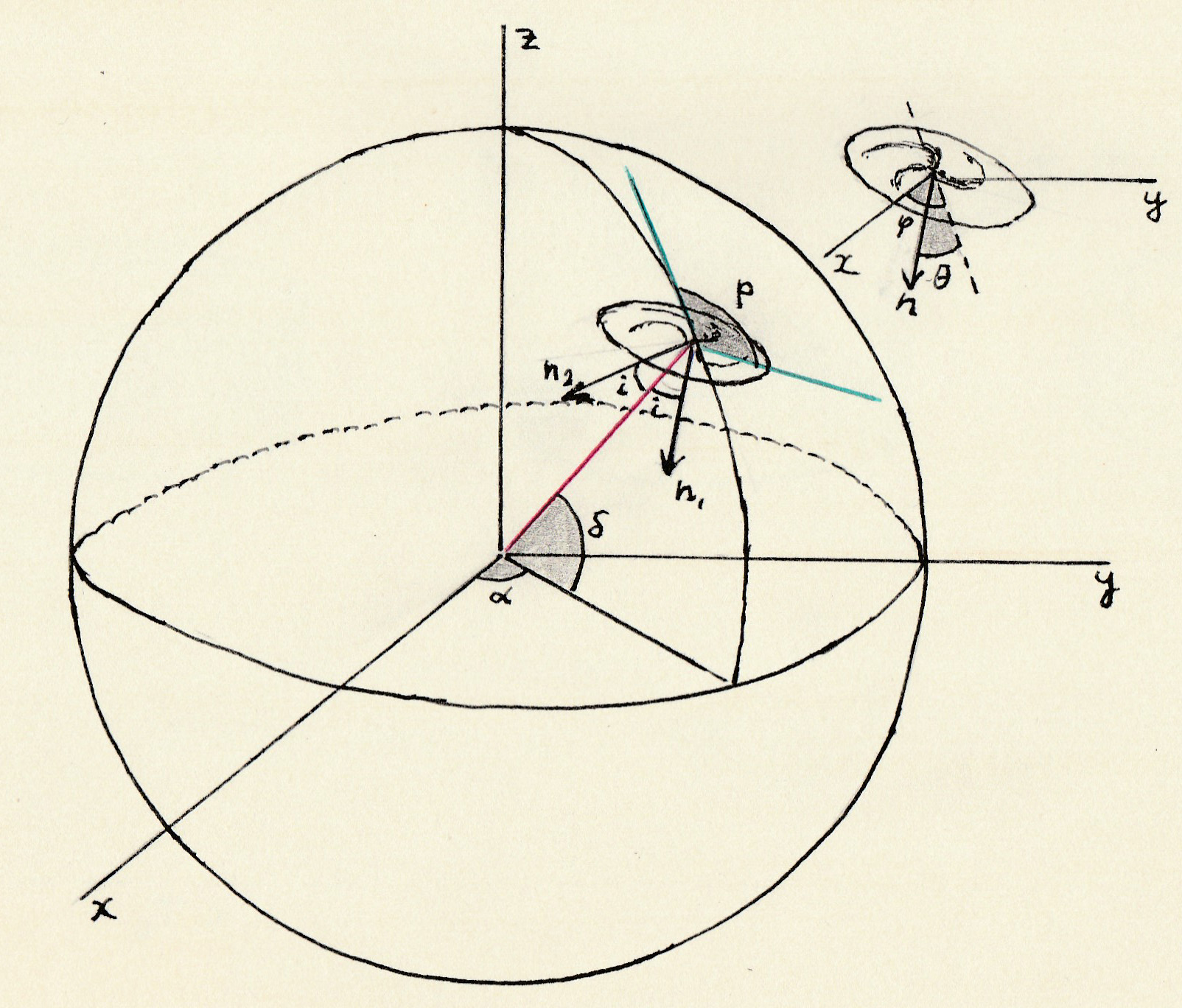}}
\caption{{\it Visualization of the spin vector of a quasar as determined by the azimuthal angle $\varphi$ and polar angle $\theta$.}}
\end{figure}
In order to obtain the 3-dimensional orientation of the spin vector (SV), one calculates the inclination (i), the azimuthal angle ($\varphi$) and polar angle ($\theta$) by the relations (Flin, et al., 1986; Pajowska, et al., 2019 )
\begin{figure}
\centerline{\includegraphics[scale=.40]{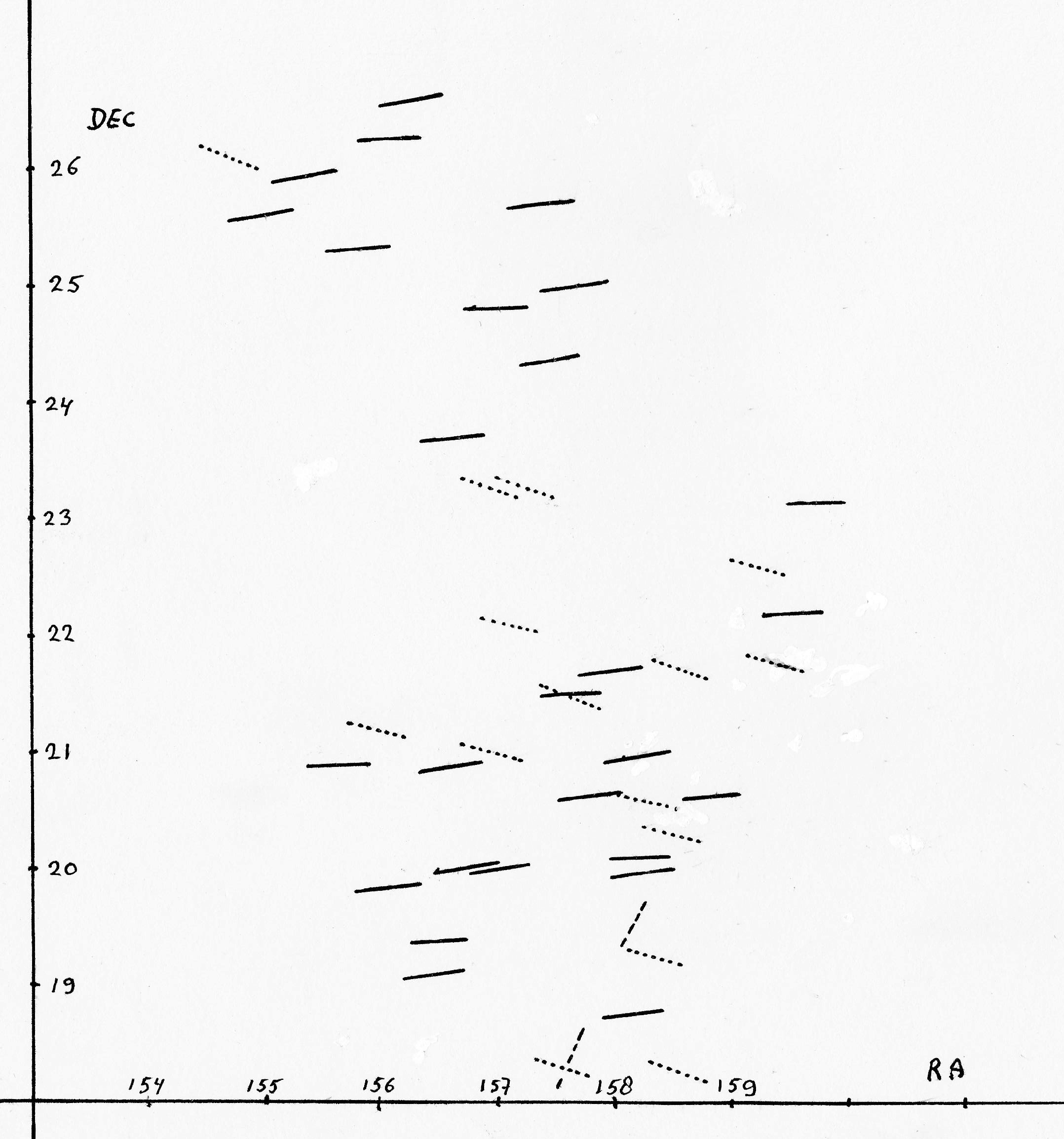}}
\centerline{\includegraphics[scale=.50]{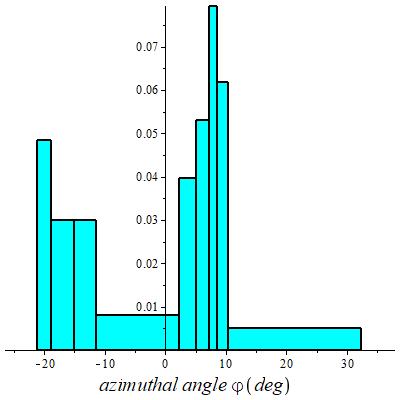}}
\caption{{\it Distribution of the azimuthal angle in LQG-18 with average redshift $\bar z = 1.49$. $N=45$.}}
\end{figure}
\begin{figure}
\centerline{\includegraphics[scale=.3]{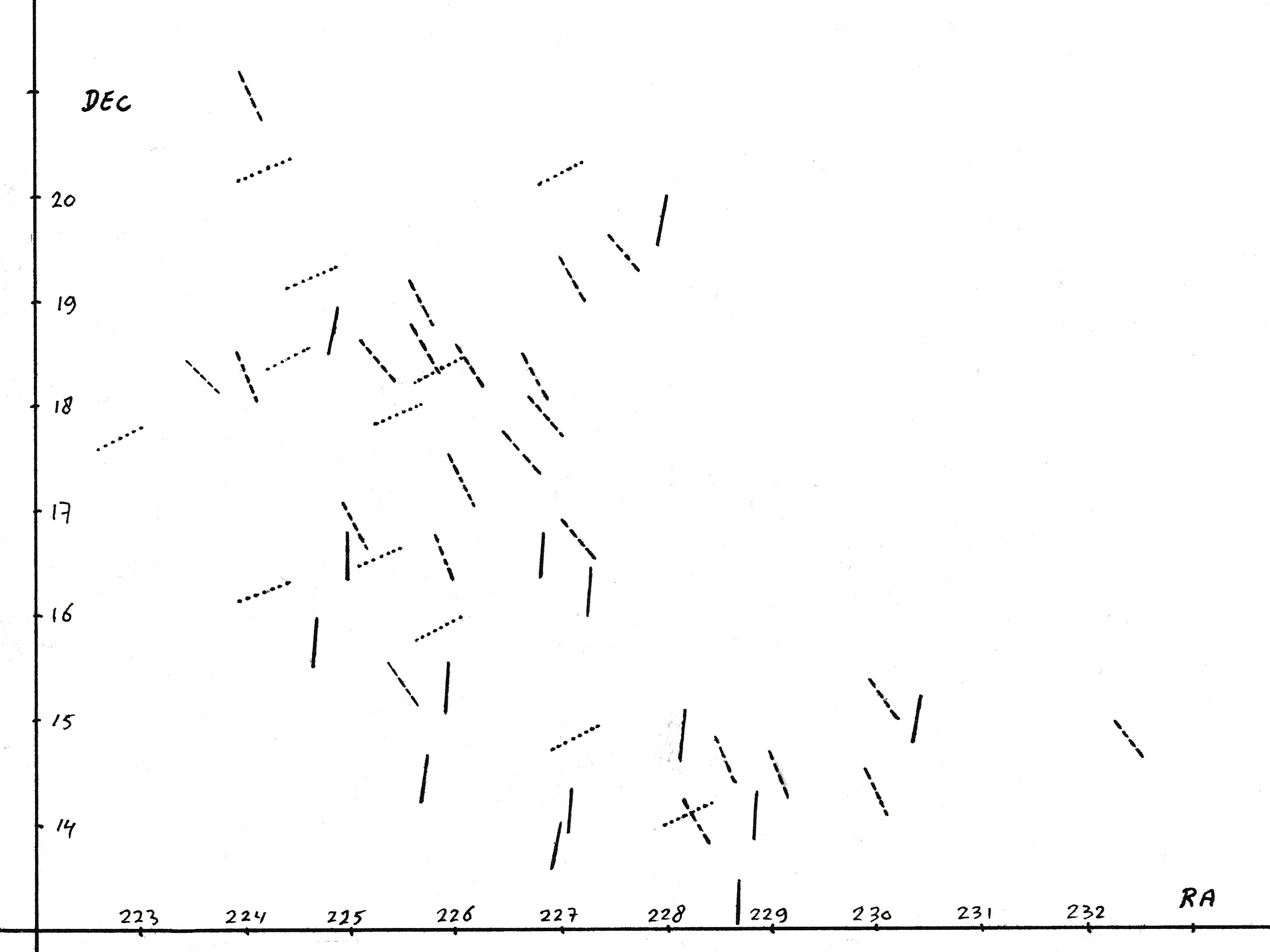}}
\centerline{\includegraphics[scale=.50]{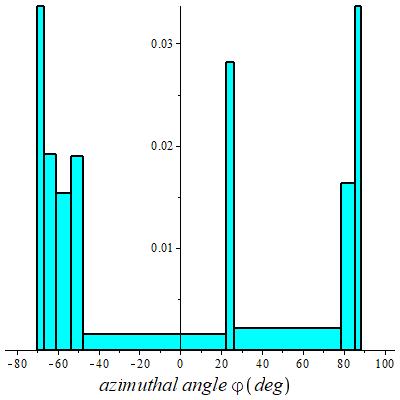}}
\caption{{\it Distribution of the azimuthal angle in LQG-12. $\bar z = 1.06$. $N=51$.}}
\end{figure}
\begin{figure}
\centerline{\includegraphics[scale=.27]{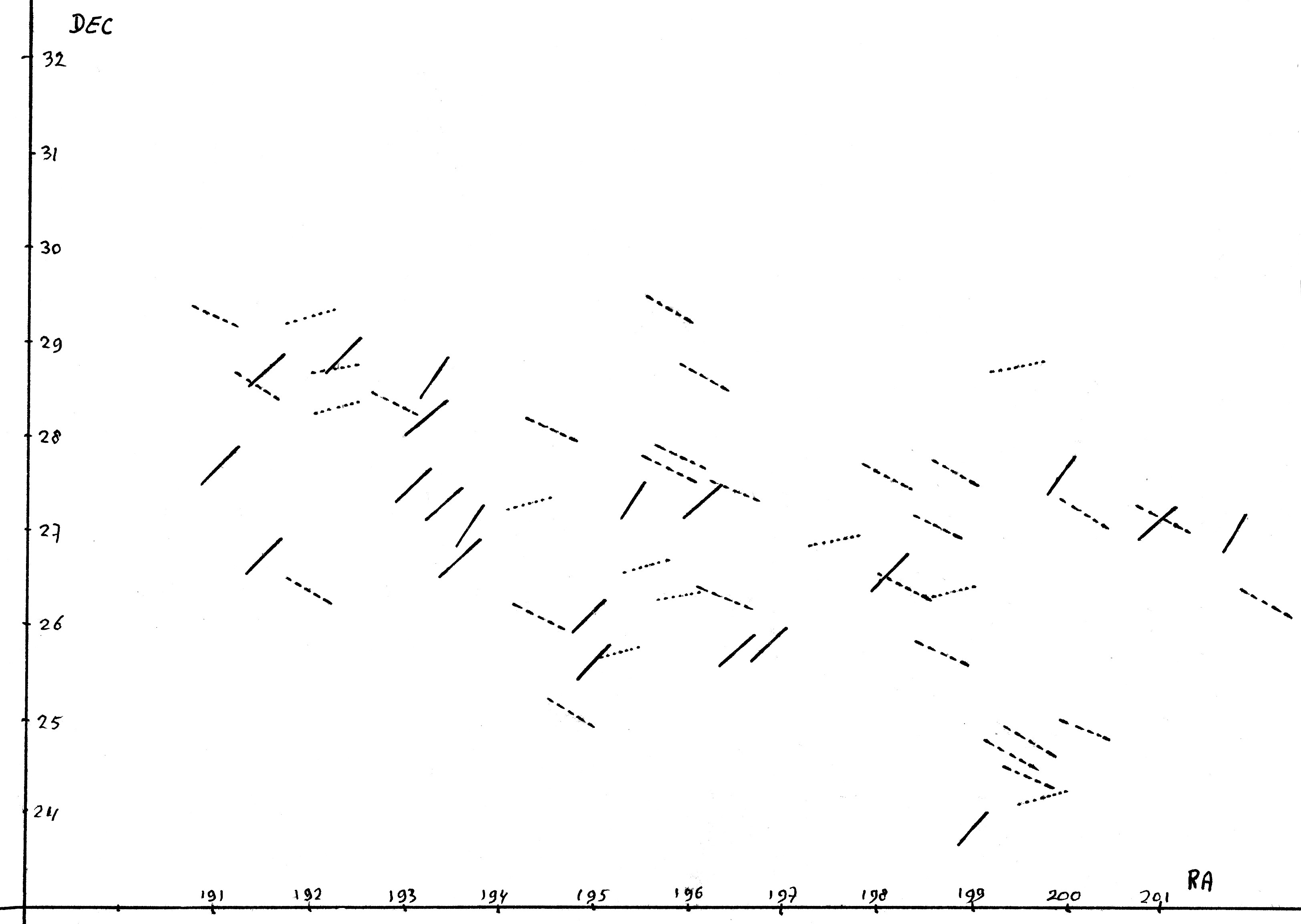}}
\centerline{\includegraphics[scale=.50]{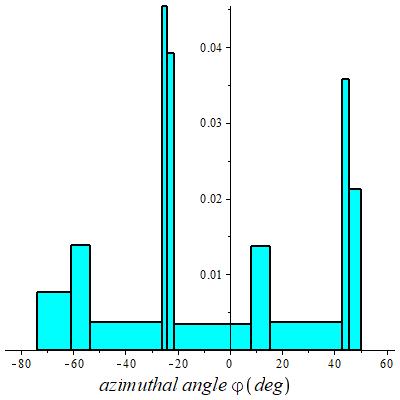}}
\caption{{\it Distribution of the azimuthal angle in LQG-4. $\bar z =1.55$. $N=62$.}}
\end{figure}
\begin{figure}
\centerline{\includegraphics[scale=.30]{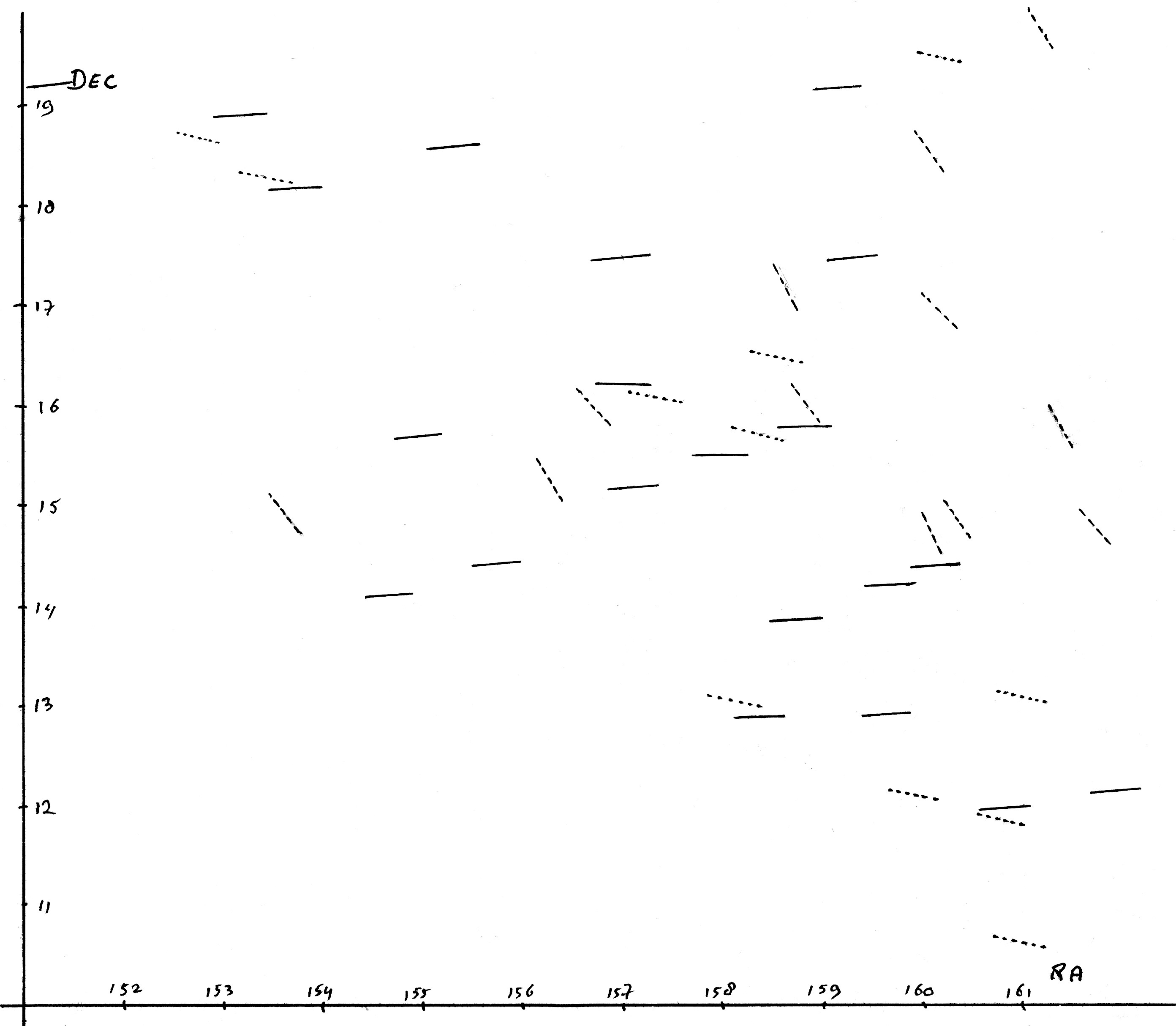}}
\centerline{\includegraphics[scale=.50]{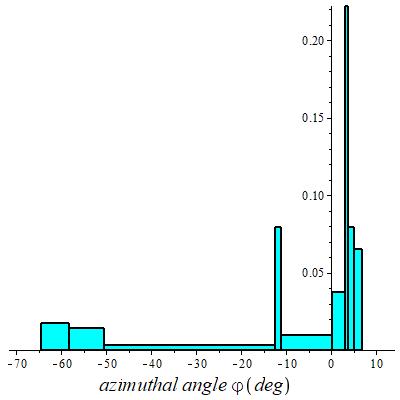}}
\caption{{\it Distribution of the azimuthal angle in LQG-19. $\bar z = 0.74$. $N=45$.}}
\end{figure} 
\begin{equation}
\cos^2{i}=\frac{\epsilon^2-\epsilon_0^2}{1-\epsilon_0^2},\label{eqn10}
\end{equation}
\begin{equation} 
\sin\theta =-\cos i \sin \delta\pm\sin i\sin p \cos \delta,\label{eqn11}
\end{equation}
\begin{equation}
\sin\varphi =\frac{-\cos i\cos \delta \sin \alpha +\sin i (\mp \sin p \sin \delta\sin\alpha \mp \cos p\cos\alpha}{\cos\theta}\label{eqn12}
\end{equation}
and
\begin{equation}
\cos\varphi=\frac{-\cos i\cos \delta \cos \alpha +\sin i (\mp \sin p \sin \delta\cos\alpha \pm \cos p\sin\alpha}{\cos\theta}.\label{eqn13}
\end{equation}
We used for the intrinsic flatness ($\epsilon_0$) the value $0.2$, which is the standard value when we have no information about  the morphological types of the analyzed galaxies (Godlowski, 2011).
By solving the equations Eq.(\ref{eqn11})-Eq.(\ref{eqn13}), one obtains four solutions for every quasar SV, $(\theta_1,\varphi_1), (\theta_2,\varphi_2), (-\theta_1,\varphi_3), (-\theta_2,\varphi_4)$, where $\varphi_3 =\varphi_1 +\pi$ and $\varphi_4=\varphi_2+\pi$. Because we don't know the direction of the galaxy rotation, we are left with only two distinct  values for the azimuthal angle in the interval  $[-\frac{1}{2}\pi, \frac{1}{2}\pi ]$. It turns out that the use of Eq.(\ref{eqn11}) and Eq.(\ref{eqn13}) is then sufficient.
If we want to make a plot of the distribution of the azimuthal angles in a specific quasar group, one can simply count for the two values for $\varphi$ (or even four, Aryal, et al., 2008), which doubles the number of galaxies. Here we don't count for the two possible orientations. We will select one  azimuthal angle from the two possible values, using the conjecture,  that there will be two peaks with different height (see theory section 2). 
In the figures 5-8 we plotted the orientations and the histograms for the  six  different redshifts.  In Figure 9 and 10  we plotted the histogram of quasar group 17 and 15. These are clear examples of the  appearance of two peaks with a second order peak around mod$(\frac{k}{6}\pi)$.
\footnote{The interested reader can obtain the tables with the calculated data from the authors.}

Without statistical analysis, it is evident that in all the  six  LQG there are preferable azimuthal directions with different peaks.

In other studies on galactic angular momentum distribution (see for example Flin, et al., 1986), in large groups of galaxies such as the Local Supercluster (Godlowski, 1993) or Abell clusters (Yadal, et al., 2017), one applies  a statistical analysis in order to exclude systematic errors.
The expected isotropic distribution curves  of the SV are obtained by performing random simulates. 
These clusters have a large value of velocity dispersion.

It should be noted, however, that we consider here rather small quasar groups with their host galaxy in a narrow RA-Dec band and at a constant redshift.(Park, et al., 2015). In each group, the distribution is different, indicating a non-Gaussian alignment  effect and not a contamination in the data. If one uses only the position angle in these groups, then the alignment is significantly less profound
(Hutsemekers, et al., 2014) and additional statistical test are again necessary. 
  
It is conjectured by the theoretical explanation of section 2, that the peaks in the azimuthal angle distribution are out of phase, and are determined by trigonometrical functions $\sin(n_i-n_j)\varphi$ and   $\cos(n_i-n_j)\varphi$ in the successive terms of ${^{4}T}_{t\varphi}$ and ${^{4}T}_{\varphi\varphi}$.  $n_i$ are the multiplicities of subsequent perturbation terms of the scalar field.
The next task is to determine  the peak heights and to compare these peaks with the theoretical prediction. More accurate data will then be necessary for high redshift.
\begin{figure}
\centerline{\includegraphics[scale=.5]{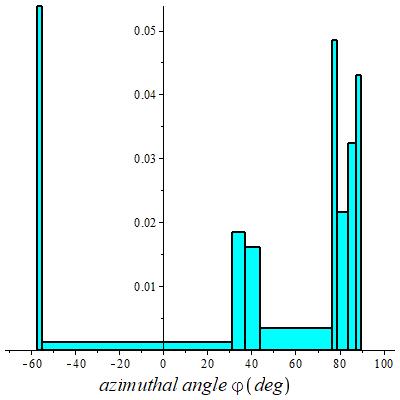}}
\caption{{\it Plot of the LQG-17. Example of the two peaks and a second order peak around $mod(\frac{k}{6}\pi)$. $\bar z =1.35$.}}
\end{figure}
\begin{figure}
\centerline{\includegraphics[scale=.5]{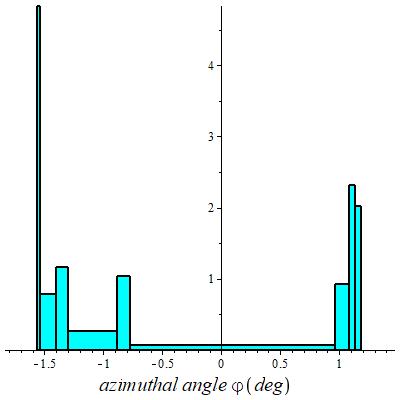}}
\caption{{\it As figure 9, now for LQG-15. $\bar z =0.96$}}
\end{figure}

\section{Conclusions}
It is found that the azimuthal angle of the spin vector of quasars in their host galaxies in the six  quasar groups under consideration, show preferred directions. This could  be explained  by an emergent azimuthal angle dependency of the Nielsen-Olesen vortices just after the symmetry breaking at GUT scale.
This is not uncommon. The inflationary period in the history of the expansion of our universe, for example, is triggered by the same field configuration at the same epoch. 
Recently, an intrinsic galaxy alignment from angular dependent primordial non-Gaussianity from massive non-zero spin quantum fields during inflation was found (Kogai, et al., 2018; Arkani-Hamad, et al., 2015).
A throughout investigation of more LQG at higher redshift will be  necessary, in order to confirm the trigonometrical distribution of the azimuthal angle of the several orders in the approximation. 
If one investigates large groups, one should carry out statistical test, in order to exclude systematic errors in the data. 
These issues are currently under investigation by the authors.
\section{Data availability}
The data underlying this article will be shared on  request to the corresponding author.
email: info@asfyon.com

\appendix
\section{The data generation}
The calculation of the two angles $\theta,\varphi$ of the host galaxies of the  five quasar group data, i.e., right ascension ($\delta$), declination ($\alpha$), eccentricity ($\epsilon=a/b$;$ a$, $b$ the major and minor  axes of the ellipse), and position angle (p), can be done with a short Maple program. The quasar data for LQG19, for example, are written in the file LQG19. It is a (N,5) matrix, with N the number of quasars in the group. Note that we work in Maple in radians.\\ \\
\begin{mdframed}[leftmargin=10pt,rightmargin=10pt]
H := Array(readdata("C:/LQG19.txt", 1, 2, 3, 4, 5));\\
M := Array(1 .. 45);\\
for i to 45 do M[i] := H(i, 4)/H(i, 3) od;\\
K := Array(1 .. 45);\\
L1 := Array(1 .. 45);L2 := Array(1 .. 45);\\
for i to 45 do K[i] := cos(e) = $\sqrt{(M[i]^2-0.04))/.96}$; \\
L1[i] := solve(K[i], e) od;
for i to 45 \\ do K[i] := cos(e) = -$\sqrt{(M[i]^2-0.04))/.96}$;\\ 
L2[i] := solve(K[i], e) od;
R1 := Array(1 .. 45);\\
R2 := Array(1 .. 45);
S1 := Array(1 .. 45);\\
S2 := Array(1 .. 45);\\
for i to 45 do R1[i] := $\{$sin(phi)\\
= (-cos(L1[i])*cos(H[i, 2]*(pi/180))*sin(H[i,1]*(pi/180))\\
+sin(L1[i])*(-sin(H[i,5]*(pi/180))*sin(H[i, 2]*(pi/180))*sin(H[i,1]*(pi/180))\\
-cos(H[i,5]*(pi/180))*cos(H[i,1]*(pi/180))))/cos(theta),\\
sin(theta) = -cos(L1[i])*sin(H[i, 2]*(pi/180))+ \\
sin(L1[i])*sin(H[i,5]*(pi/180))*cos(H[i,2]*(pi/180))$\}$;\\ R2[i] := solve(R1[i], $\{$phi, theta$\}$ od\\
for i to 45 do S1[i] := $\{$sin(phi) = \\\ 
(-cos(L2[i])*cos(H[i,2]*(pi/180))*sin(H[i,1]*(pi/180))\\
+sin(L2[i])*(-sin(H[i,5]*(pi/180))*sin(H[i,2]*(pi/180))\\*sin(H[i,1]*(pi/180))\\
-cos(H[i,5]*(pi/180))*cos(H[i,1]*(pi/180))))/cos(theta),\\
sin(theta) =-cos(L2[i])*sin(H[i,2]*(pi/180))\\ 
+sin(L2[i])*sin(H[i,5]*(pi/180))*cos(H[i,2]*(pi/180))$\}$;\\
S2[i] := solve(R1[i], $\{$phi, theta$\}$ od\\
\end{mdframed}
\bsp	
\label{lastpage}

\begin{thebibliography}{30}
\bibitem{abr}
Abrikosov, A. A. (1957) {\it Soviet Physics JETP} {\bf 5}, 1774, 1957.
\bibitem{agar}
Agarwal, N. , Kamal, A. and Jain, A. (2011)  {\it Phys. Rev. D} {\bf 83}, 065014. 
doi.org/10.1103/PhysRevD.83.065014
\bibitem{and}
Anderson, M.R. (2003) The Mathematical Theory of Cosmic Strings. IoP publishing, Bistol, UK.
\bibitem{ark}
Arkani-Hamed, N. and Maldacena, J. (2015)  arXiv: hep-th/1503.08043
\bibitem{aryal}
Aryal, B., Kafle, P. R. and Saurer, W. (2008) {\it Mon. Not. R. astr. Soc.} {\bf 389}, 741.
doi.org/10.1111/j.1365-2966.2008.13494.x
\bibitem{bog}
Bogomol'nyi, E. (1976) {\it Sov. J. Nucl. Phys.} {\bf 24}, 449. doi.org/10.12691/amp-2-3-8
\bibitem{choc1}
Choquet-Bruhat, Y. (1969) {\it Commun. Math. Phys.} {\bf 12}, 16.
\bibitem{choc2}
Choquet-Bruhat, Y. (1977) {\it Gen. Rel. Grav.} {\bf 8}, 561.
\bibitem{ono}
D'Onofrio, M., Marziani, P., Sulentic, J. W. (2012) Fifty Years of Quasars. Springer, Berlin, Germany. doi.org/10.1007/978-3-642-27564-7
\bibitem[Felsager(1987)]{fels}
Felsager, B. (1987)  Geometry, particles and fields. Odense University press: Odense, Denmark.
\bibitem{flin}
Flin, P. and Godlowski, W. (1986) {\it Mon. Not. R. astr. Soc.} {\bf 222}, 525. 
doi.org/10.1093/mnras/235.3.857
\bibitem{garf}
Garfinkle, D. (1985)  {\it Phys. Rev.} {\bf D32}, 1323. doi.org/10.1103/PhysRevD.32.1323
\bibitem{gins}
Ginzburg, V. L. and Landau, L. D. (1950) {\it Zh. Eksp. Teor. Fiz.} {\bf 20}, 1064.
\bibitem{god1}
Godlowski, W. (1993) {\it Mon. Not. R. Astron. Soc} {\bf 265}, 874. doi.org/10.1093/mnras/265.4.874
\bibitem{god2}
Godlowski, W. (2011) {\it Acta Phys. Pol.} {\bf B42}, 2323. doi.org/10.5506/APhysPolB.42.2323
\bibitem{huts2}
Hutsemekers, D., Cabanac, R., Lamy, H. and Sluse, D. (2005) {\it Astron. Astyrophys.} {\bf 441}, 915. doi.org/10.1051/0004-6361:20042163
\bibitem{huts1}
Hutsemekers, D., Braibant, L., Pelgrims, V. and Sluse, D. (2014)  {\it Astron. Astrophys.}
{\bf 572}, A18.  doi.org/10.1051/0004-6361/201424631 
\bibitem{kogai}
Kogai, K., Matsubara, T., Nishizawa, A. J. and Urakawa, Y. (2018) {\it J. Cosm.  Astroparticles} {\bf 2018} doi.org/10.1088/1475-7516/2018/08/014
\bibitem{lebov}
Lebovitz, N. R. (1967) Bifurcation and stability problems in astrophysics, in Applications of bifurcation theory. Academic Press, New York, USA
\bibitem{mal}
Malla, J. R., Aryal, B. and Saurer, W. (2019) {\it NUTA Journ.} {\bf 6}, 12.
\bibitem{mant}
Manton, N. and Sutcliffe, P. (2007)  Topological solitons. Cambridge University press, Cambrigde,  UK. 
\bibitem{niels}
Nielsen, H.B. and Olesen, P. (1973) {\it Nucl. Phys.} {\bf B61}, 45.
\bibitem{paj}
Pajowska, P., Godlowski, W., Zhu, Z., Poliela, J. and Panko, E. (2019) {\it J. Cosm. Astroparticles} {\bf 02}, 005. doi.org/10.1088/1475-7516/2019/02/005
\bibitem{park}
Park, C., Song, H., Einasto, M., Lietzen, H. and Heinamaki, P. (2015) {\it J. Korean Astr. Soc.} {\bf 48},75. doi.org/10.5303/JKAS.2014.00.0.1
\bibitem{pel1}
Pelgrims, V. and Hutsemekers, D. (2016) {\it Astron. and Astrophys.} {\bf 590}, A53. doi.org/10.1051/0004-6361/201526979 
\bibitem{pel2}
Pelgrims, V. (2016) arXiv: astro-ph/160405141.
\bibitem{pol}
Poltis, R. and Stojkovic, D. (2010) {it Phys. Rev Lett.} {\bf 105}, 161301. doi.org/10.1103/PhysRevLett.105.161301
\bibitem{ran1}
Randall, L. and Sundrum, R. (1999) {\it Phys. Rev. Lett.}, {\bf 83}, 3370. doi.org/10.1103/PhysRevLett.83.3370
\bibitem{ran2}
Randall, L. and Sundrum, R. (1999) {\it Phys. Rev. Lett.} {\bf 83}, 4690.doi.org/10.1103/PhysRevLett.83.4690
\bibitem{slag6}
Slagter, R. J. (2001) {\it Class. and Quantum Grav.} {\bf 18}, 463. doi.org10.1088/0264-9381/18/3/308
\bibitem{slag1}
Slagter, R.J. and Pan, S. (2016) {\it Found. of Phys.} {\bf 46}, 1075. doi.org/10.1007/s10701-016-0002-2
\bibitem{slag2}
Slagter, R.J. (2016) {\it J. of Mod.Phys.} {\bf 7, no 6}, 501. doi.org/10.4236/jmp.2016.76052 
\bibitem{slag3}
Slagter, R.J. (2017) {\it J. of Mod.Phys.} {\bf 8, no 2}, 163. doi.org/10.4236/jmp.2017.82015 
\bibitem{slag5}
Slagter, R. J. (2017) Spacetime Physics 1907-2017. Minkowski Inst. Press, Montreal, Canada.
\bibitem{slag4}
Slagter, R.J. (2018) {\it Int, J. of Mod.Phys.D} {\bf 27}, 1850094. doi.org/10.1142/S0218271818500943
\bibitem{taylor}
Taylor, A.R. and Jagannathan, P. (2016) {\it Mon. Not. Roy. Astr. Soc.} {\bf 459}, {\em 459}, L36.
doi.org/10.1093/mnrasl/slw038
\bibitem{vil}
Vilenkin, A. and Shellard, E.P.S. (1994) Cosmic Strings and Other Topological Defects. Cambridge University press, Cambrigde,  UK. 
\bibitem{yadav}
Yadav, S. N., Aryal, B. and Saurer, W. (2016)  arXiv: astro-ph/160602881.
\bibitem{yadav2}
Yadav, S. N., Aryal, B. and Saurer, W. (2017) {\it Res. in Astron. Astrophys.} {\bf 17}, 64.
\end{thebibliography}
\end{document}